\magnification=1200 
\def\th{\theta} 
\def\rf{\bar\rho} 
\def\dl{\delta} 
\def\Dl{\Delta} 
\def\ep{\epsilon} 
\def\kp{\kappa} 
\def\vx{\vec x} 
\def\vy{\vec y} 
\def\vxy{|\vx-\vy|} 
\def\cxy{{\vx-\vy\over\vxy^3}} 
\def\inte{\int^\infty_{E_0}dE\,} 
\def\ints{\int_{-y_1}^{x_e-y_1}ds\,{s\over(s^2+b^2)^{3/2}}} 

\leftline{\bf Post-Newtonian Sachs-Wolfe Effect}
\leftline{Eric V. Linder} 
\bigskip 
\leftline{Imperial College, Prince Consort Road, London SW7 2BZ, 
England} 
\leftline{\quad el@ic.ac.uk} 
\bigskip\medskip
 
\noindent{\bf Abstract.} 
Deviations from the Friedmann-Robertson-Walker (FRW) cosmological model in
the form of density inhomogeneities induce 
observational effects on the light
propagating through these fluctuations.  Using a rigorously parametrized 
metric whose pseudo-Newtonian potential is related to 
(possibly nonlinear) density inhomogeneities through a relativistic 
Green function, the behavior of radiation propagating through this 
approximation to our universe is investigated, without relying on any 
spatial averaging or late time, short perturbation wavelength Newtonian 
limit.  In certain regimes the energy shift of a photon 
due to density fluctuations, the 
Sachs-Wolfe effect, is found to deviate significantly from the 
linearized relativity (Poisson-Newton) result, but the corresponding 
cosmic microwave background temperature fluctuations are still below 
$10^{-6}$ on small scales.  On large scales this can be treated as 
an effective transfer function for the 
density power spectrum, altering the scaling of the amplitudes of large 
versus small scale power. 
\medskip 
\leftline{{\bf Key words:} Relativity, Gravitation, Cosmic microwave 
background} 
\bigskip\medskip 
 
\leftline{\bf 1. Introduction} 
\medskip\noindent 
A self consistent treatment of observations in a realistically inhomogeneous 
universe involves relativistic cosmological calculation of both
the influence of the inhomogeneity on the metric and the metric on the light. 
Often both are treated only within the 
late time, short wavelength Newtonian approximation where the
density determines a spatial metric fluctuation by the Poisson equation and
the metric influences the propagation through an essentially local impulse.
Here we use recent improvements beyond ``Newtonian'' order to
investigate relativistically 
light propagation, especially from the last scattering surface of 
the cosmic microwave background radiation (CMB).  

In \S2 we use a 
relativistically rigorous metric (Futamase 1989) describing a universe 
deviating by gravitational potential perturbations 
from FRW, accurate to a well defined order, to obtain the 
Einstein field equations.  The Green function solution of Jacobs, Linder, 
\& Wagoner (1992=JLW1; 1993=JLW2) relates the gravitational potential to 
the density inhomogeneity for arbitrary density contrasts, i.e. without 
restrictions to the linear regime.  We identify those regions in parameter
space where post-Newtonian effects are appreciable, as well as deriving 
analytic expressions for the derivatives of the potential, useful in 
light propagation calculations.  In \S3 we concentrate on the 
Sachs-Wolfe effect, where density inhomogeneities generate temperature 
anisotropies in the CMB, evaluating the deviation of the Green function 
results from the standard Newtonian ones.  The results are given both 
numerically and by analytic order of magnitude arguments to reveal where 
the deviations arise. 
In addition we find an effective 
correction to the density power spectrum $P_k$ which alters the 
intrinsic scaling of its amplitude from small to large scales. 

While the redshift of light propagating through inhomogeneities is 
well studied (see, e.g.,  Ma \& Bertschinger 1995 for the Sachs-Wolfe 
effect in linear perturbation theory and Seljak 1996 for the 
Rees-Sciama effect in numerical simulations), it has been considered 
within linearized general relativity and not the post-Newtonian formalism 
of Futamase 1989, JLW1, and JLW2.  That possibly significant 
differences may arise can be seen from the diffusion equation analogy 
of JLW2 as well as the relativistic approaches of Kodama \& Sasaki 
1984 and Futamase \& Schutz 1983, which agree with the results here 
that there exist corrections to the linearized general relativistic 
solution -- the Poisson-Newton equation between the gravitational 
potential perturbation appearing in the metric, $\phi$, and the 
density fluctuation $\delta\rho$.  That is just the late time, 
short wavelength ``Newtonian'' approximation. 

\bigskip 
\leftline{\bf 2. Gravitational potential}
\medskip\noindent 
Under the assumptions that the gravitational potential fluctuations $\phi$
are small, para\-metrized by $\ep^2\ll1$, and their peculiar accelerations 
$\nabla\phi\sim\ep^2/\kp$ are small (thus ensuring that peculiar motions 
remain 
much less than the speed of light) for characteristic inhomogeneity scales
$l=\kp L$ with $L$ the background curvature or horizon length scale, the
following metric provides a consistent description: 
$$ds^2=a^2(\eta)\,\bigl[-(1+2\phi)d\eta^2+(1-2\phi)\gamma_{ij}dx^i
dx^j\bigr]\quad,\eqno(1)$$ 
(Futamase 1989; JLW1).  Here $a$ is the FRW
expansion factor (to the order required), $\eta$ the conformal time, 
and $\gamma_{ij}$ the spatial part of the conformally
stationary Robertson-Walker metric.  The conditions are expected to hold
cosmologically everywhere far outside the radii of black
holes and neutron stars.  We call the full time dependence and presence 
of $\phi$ in the spatial part of the metric the post-Newtonian corrections. 

The Einstein field equations produce the following relation between the
scalar harmonic modes of $\phi$ and the density contrast $\Dl=\dl\rho/\rf$, 
$$3{\dot a\over a}\dot\phi(\eta,\vec q)+(q^2+8\pi a^2\rf-6k)\,\phi(\eta,\vec
q) =-4\pi a^2\rf\Dl(\eta,\vec q),\eqno(2)$$ 
(JLW1; JLW2).  Here $k$ is the trichotomic
FRW curvature parameter, $\bar\rho$ the unperturbed density, 
and $\vec q$ is the mode variable conjugate to
position.  No restrictions are made on the size of $\Dl$, i.e. the 
density field could be nonlinear. 

Solution of this equation gives (JLW2) 
$$\eqalign{\phi(\eta,\vec x)=-(4\pi/3)\,\int_{\eta_0}^\eta du\,&(a^3
\rf/ \dot a)\int d^3\vec y\,G(u,\eta,\vx,\vy)\,\Dl(u,\vy)\cr 
&+\int d^3\vec y\,G(\eta_0,\eta,\vx,\vy)\,\phi(\eta_0,\vy),\cr}\eqno(3)$$ 
with the Green function in the $k=0$ case (used here throughout) 
$$\eqalign{G(u,\eta,\vx,\vy)&=[a(u)/a(\eta)]\,[4\pi C(u,\eta)]^{-3/2}
\exp\{-\vxy^2/4C(u,\eta)\}\cr 
C(u,\eta)&=(1/3)\int_u^\eta dw\,(a/\dot a)\quad.\cr}\eqno(4)$$ 
Transformation of the time integration variable to $E=\vxy/2C^{1/2}(u, \eta)$
reveals 
$$\phi(\eta,\vx)=-(2/\sqrt{\pi})\,\int d^3\vy\,a^{-1}(\eta)\vxy^{-1}
\int_{E_0}^\infty dE\,[a^3\rf\Dl(E,\vy)]\,e^{-E^2},\eqno(5)$$ 
where $E_0=\vxy/2C^{1/2}(\eta_0,\eta)$.  

In the Newtonian limit of late times
($\eta\gg\eta_0$) and subhorizon scales ($\kp\ll1$), $E_0\to0$ and 
it is found that 
$$\phi(\eta,\vx)=-\int d^3\vy\,a^3{\rf\Dl(\eta,\vy)\over a\vxy}\eqno(6)$$ 
(JLW2), the usual result.  [We have neglected the second, initial 
conditions term from (3) due to its exponential die off far from the 
initial hypersurface, but see \S3.3.] 

Going beyond JLW2 we adopt 
a dust background ($a=a_0\eta^2=2H_0^{-1}\eta^2$ where $H_0$
is the Hubble constant) and inhomogeneity behavior $\Dl(u,\vy)=f(\vy)
\rf^{-1}\,[a(u)/a_0]^{-3+p}$ to yield 
$$\phi(\eta,\vx)=-{2\over a_0\sqrt{\pi}}\,\left[{a(\eta)\over a_0}\right]^
{p-1}\int 
d^3\vy\,{f(\vy)\over\vxy}\,\inte \left[1-3{\vxy^2\over\eta^2}E^{-2}
\right]^p e^{-E^2}.\eqno(7)$$ 
In the Newtonian limit this reduces to 
$\phi(\eta,\vx)=-a_0^{-1}[a(\eta)/a_0]^{p-1}\int d^3\vy\,f(\vy)/\vxy$. 
Some particular cases of interest are $p=0$ (density unevolving in physical
coordinates) and $p=1$ (as for the growth of linear density fluctuations
$\Dl\ll1$): 
$$\eqalignno{\phi(\eta,\vx)_{p=0}&=-{1\over a(\eta)}\int d^3\vy {f(\vec
y)\over\vxy}\,{\rm erfc}(E_0)&(8a)\cr  \phi(\eta,\vx)_{p=1}&= 
-{1\over a_0}\int d^3\vy {f(\vy)\over\vxy}\left\{{\rm erfc}(E_0)\Bigl[1+6
{\vxy^2\over\eta^2}\Bigr]-{2\over\sqrt{\pi}}\Bigl(1-{\eta_0^2\over\eta^2}
\Bigr)E_0  e^{-E_0^2}\right\}&(8b)\cr}$$ 
with erfc the complementary error function. 

Since the geodesic equations determining light propagation involve
derivatives of the potential we calculate in comoving coordinates 
$\vec\nabla\phi$ and $\nabla_i\nabla_j\phi$. 
$$\eqalign{\vec\nabla\phi&=-{2\over a_0\sqrt{\pi}}\Bigl[{a(\eta)\over
a_0}\Bigr]^{p-1}\int d^3\vy\,f(\vy)\cxy\,I\cr 
I&=\inte U^pe^{-E^2}\,+\,
6p{\vxy^2\over\eta^2}\inte U^{p-1}E^{-2}e^{-E^2}\,+\, U_0^pE_0e^{-E_0^2}\cr 
U&=1-3{\vxy^2\over\eta^2}E^{-2},\cr}\eqno(9)$$ 
with special cases 
$$\eqalignno{\vec\nabla\phi_{p=0}&={1\over a(\eta)}\int d^3\vy\,f(\vy)\cxy 
\,\Biggl[{\rm erfc}(E_0)+{2\over\sqrt{\pi}}E_0e^{-E_0^2}\Biggr]&(10a)\cr 
\vec\nabla\phi_{p=1}&={1\over a_0}\int d^3\vy\,f(\vy)\cxy\,\Biggl[
\Bigl(1-6{\vxy^2\over\eta^2}\Bigr)\,{\rm erfc}(E_0)\,+\,{2\over
\sqrt{\pi}}E_0e^{-E_0^2}\Biggr].&(10b)\cr}$$ 
The tidal field is $\nabla_i\nabla_j\phi$, necessary for calculations
involving shear of a light ray bundle and the resulting image distortions.
Because of its length we do not show the expression for it here but it is
interesting to consider the Laplacian $\nabla^2\phi$.  For $p=0$ 
$$\eqalign{\nabla^2\phi&=4\pi a^{-1}\left\{f(\vx)-\pi^{-3/2}\int d^3\vy\,
{f(y)\over\vxy^3} E_0^3e^{-E_0^2}\right\}\cr 
&=4\pi a^{-1}\left\{f(\vx)-(4\pi C_0)^{-3/2}\int d^3\vy\,
f(\vy)\,e^{-\vxy^2/4C_0}\right\},\cr}\eqno(11)$$ 
where $C_0=C(\eta_0,\eta)$. 

The second term thus illustrates the (not linearized) general 
relativistic correction to the
Poisson equation (in the fully specified longitudinal gauge), 
involving a gaussian weighting over the extent of the
inhomogeneity.  In the ``Newtonian'' limit the fluctuation is restricted to
regions much smaller than the dispersion $(2C_0)^{1/2}$ 
so its contrast is averaged to zero,
leaving the Poisson result, while in the opposite (``superhorizon''
fluctuation) limit the gaussian becomes a delta function, giving the Laplace 
equation appropriate for a uniform density field. 
\bigskip 
\leftline{\bf 3. Sachs-Wolfe effect}
\medskip\noindent
Given the metric (1) and the results (7), (9), (11) 
one can investigate light propagation behavior by
writing down the geodesic equation, geodesic deviation, and the beam or
Raychaudhuri equation.  Properties of the photon bundle, such as
convergence and shear, and applications of the geodesic deviation equation to
astrophysical problems such as correlations of observables in terms of
the density power spectrum are in ongoing research. 
Here we concentrate on the geodesic equation for individual photon 
four momentum, in particular the Sachs-Wolfe effect on the redshift. 

In longitudinal gauge ($g_{0i}=0$ and $g_{ij}$ proportional to $\gamma_{ij}$; 
see Bardeen 1980 and Kodama \& Sasaki 1984 for gauges and gauge 
invariance) the expression for the frequency of a photon emitted 
at $\eta_e$ becomes 
$$k^0(\eta)=k^0(\eta_{e})\,[a(\eta_{e})/a(\eta)]\,\Bigl(1-2\int_
{\eta_{e}}^\eta du\,\hat n\cdot\vec\nabla\phi\Bigr)\quad.\eqno(12)$$ 
(Note the more familiar time derivative of $\phi$ occurs in the unfully 
specified synchronous gauge expression, although one could convert 
the gradient into a time derivative and a surface term.) 
The ratio of expansion factors is the background cosmological redshift and
$\hat n$ is the photon propagation unit vector.  The 
inhomogeneity induced redshift, i.e. the Sachs-Wolfe effect, is then 
$$z=2\int_{\eta_{e}}^1 d\eta\,\hat n\cdot\vec\nabla\phi\quad.\eqno(13)$$ 

Upon adopting a density field $\Dl(u,\vy)$ one can compute the 
gravitational potential gradient by (9) and hence obtain the redshift, or 
equivalently temperature anisotropy in the CMB.  This, being an observable, 
is gauge independent.  We consider compact density distributions, both 
static and time dependent, and then a field of inhomogeneities, along with 
the Fourier representation. 
\bigskip 
\leftline{\bf 3.1 Static point mass} 
\medskip\noindent 
To derive analytic estimates of the dependence of the redshift on density
parameters we begin by adopting the model of a simple point inhomogeneity of
mass $m$ at comoving position $\vy=(y\cos\th,y\sin\th,0)$.  The $x$-axis is
aligned with the light ray (perturbations to the photon path produce effects 
of smaller order than we
consider) and $y=1-(1+z_m)^{-1/2}$ where $z_m$ is the cosmological redshift
of the mass.  The density contrast is $\Dl(u,\vx)=m\rf^{-1}a^{-3}(u)\,\dl^
{(3)}(\vx-\vy)$, corresponding to $p=0$.  Such a static case is intended 
as a toy model only  and is not  realistic   over long times or large 
scales. 

For analytic computation rewrite (13) with (10a) as 
$$z=2{m\over a_0}\ints (1-y_1-s)^{-2}\left[{\rm erfc}(E)+(2/\sqrt{\pi})
Ee^{-E^2}\right],\eqno(14)$$ 
where the photon is emitted at $x_e=1-\eta_{e}=1-(1+z_e)^{-1/2}$, 
$y_1=y\cos\th$, $s=x-y_1$, and $s=0$ corresponds to closest 
approach at impact parameter $b=y\sin\th$.  We have suppressed the 
subscript $0$ on $E$.  First consider the Newtonian case, where the bracketed
quantity is unity.  Around $s=0$ the integrand is nearly odd so the dominant
term from the symmetric part of the interval requires expanding
$(1-y_1-s)^{-2}$ to obtain an overall term even in $s$.  One readily sees
that this gives a logarithmic integral $\sim\ln b^{-1}$.  (Although  the 
mass is static, the background is not so an energy shift is  expected; 
but see the next subsection for the $p=1$ case.) 

Another interval giving a
potentially large  contribution is when $s\approx x_e-y_1$ or $x\approx x_e$.
Here the integral varies as $(1-x_e)^{-1}=(1+z_e)^{1/2}$.  Physically this
corresponds to the early epoch when the universe was smaller and so the 
mass was nearer the source, or
equivalently the mass per physical volume was greater.  The final order of
magnitude estimate is therefore $z\approx (m/a_0)[(1+z_e)^{1/2}+\ln b^{-1}]$.
For small angles $b\approx y\th$.  In terms of our original parametrization
$m/a_0\sim\ep^2\kp$ and $\th\ge\kp$ so 
the ``Newtonian'' result is $z_N\le \ep^2\kp[(1+z_e)^{1/2}+\ln
\kp^{-1}]$.  Recall that $m/a_0\approx10^{-8}(m/10^{15}M_\odot)$, 
$10''\approx5\times10^{-5}$ radians so $\ln\kp^{-1}\approx10$, and the
microwave background, for example, originates from $z_e\approx10^3$ so 
$z_N={\cal O}(100\,mH_0<10^{-6})$. 

In the post-Newtonian case, near the mass the integrand is 
approximately the same, the
correction factor being $1+{\cal O}(E^3)\approx1+{\cal O}(\th^3)$, while far
away the Green function cuts off the contribution exponentially so
$z_{PN}\sim mH_0\ln\th^{-1}$ only.  The top half of Table 1 
gives $z$ in units of $mH_0$ for the static case, 
showing that the analytic dependences hold well. 

\bigskip
\centerline{\bf Table 1:\quad Point Mass Sachs-Wolfe Effect} 
\centerline{$(z/mH_0)_{N/PN}$} 
\medskip 
\settabs\+\hskip1truein &M=a0\qquad\qquad\qquad&M=100n\qquad\qquad 
&M=100n\qquad\qquad &\cr 
\+&{\bf $\Delta\sim a^0$}&$\theta=1''$&$\theta=10''$&$\theta=100''$\cr 
\+&$z_m=1$&218/134&192/108&166/82\cr 
\vskip-4pt
\+&$z_m=3$&527/348&453/275&380/201\cr 
\medskip 
\+&{\bf $\Delta\sim a$}&&&\cr 
\+&$z_m=1$&&1.93/-1.51&\cr 
\vskip-4pt
\+&$z_m=3$&&-0.14/-3.64&\cr 

\bigskip 
\leftline{\bf 3.2 Time varying point mass} 
\medskip\noindent 
We now examine $m=m_0(a/a_0)^p$, corresponding to an accreting (or
evanescing) mass.  In density perturbation theory inhomogeneities grow due to
gravitational instability as $a^1$ in a $k=0$ dust background.  Here of
course we are not restricted to linear fluctuations $\Dl\ll1$; the
parametrization, as can be seen from (2) or as discussed in JLW2, is ${\cal
O}(\Dl)=\ep^2/\kp^2$, which can be large.  

The analog of (14) in the Newtonian limit is 
$$z_N=2{m\over a_0}\ints (1-y_1-s)^{2(p-1)}.\eqno(15)$$ 
Again splitting up the interval we find near $s=0$ that $z_N\approx
(1-p)(m/a_0) 
\ln b^{-1}$.  As before, the other possibly significant contribution 
arises near $x=x_e$ (for $p<1/2$) and is found to vary as $(1-x_e)^{2p-1}=
(1+z_e)^{(1-2p)/2}$.  Thus the total is $z_N\approx mH_0[(1-p)\ln \th^{-1}+ 
(1+z_e)^{(1-2p)/2}]$.  This of course agrees when $p=0$ with the static 
result, since $\th\sim b\sim\kp$.  Inclusion of the post-Newtonian 
correction again cuts out the $1+z_e$ contribution.

Note that for $p>1/2$ the logarithm dominates.  Also note that for $p=1$, as 
in linear theory, we find $z_N=2(m/a_0)[(s_{o}^2+b^2)^{-1/2}-(s_{e}^2+b^2)
^{-1/2}]$ so if $s_{o}=-s_{e}$, i.e. the observer and emitter are situated
symmetrically with respect to the inhomogeneity and so at the same value of
gravitational potential, then there is zero redshift.  Classically this
corresponds to a ball gaining kinetic energy as it rolls into a dip then
losing it coming out.  If it rises to the same height on the far side and if
the dip is time independent then its final velocity equals its initial.  We
see that for $p=1$ the Newtonian potential is time independent and only the
endpoints contribute -- the usual result. 

However, even for $p=1$ and symmetric observer-emitter
geometry there is a redshift (actually a blueshift) in the post-Newtonian 
case due to symmetry breaking by the distance dependence of the 
correction, i.e. roughly $\phi\sim 
a^0{\rm erfc}(E_0)$ which is neither symmetric nor time independent.  
The shift is of order $mH_0$,
just as for the asymmetric endpoint case.  The Sachs-Wolfe
effect for point masses varying with time as $a$ is given in 
the bottom half of Table 1. 
Because of the vanishing
of the logarithmic term the results are angle independent (for small angles).
For $p>1$ the logarithmic term reappears and the results regain the angle 
dependence of the $p=0$ case. 
The generalization to an extended matter distribution, 
e.g. a spherical density inhomogeneity, does not significantly affect the 
results (Kendall 1993). 

The time dependent effect
for a dynamically evolving isolated Newtonian density fluctuation, e.g. 
a cluster decoupled from the universal expansion, is
sometimes known as the Rees-Sciama effect (Rees and Sciama 1968).  A
gravitational 
dynamical time scale is $(\rf\Dl)^{-1/2}\sim\kp/\ep$ so we expect $z\sim \int
d\eta \partial_t\phi\sim b(\rf\Dl)^{1/2}\phi$, parametrized as $\kp\cdot
\ep/\kp\cdot\ep^2\sim\ep^3$.  Alternately consider an effective mass within 
a fixed comoving radius at the times a light ray enters and leaves the
potential well: $m(t_{out})\approx m(t_{in})\,[1+{\cal O}(v/c)]$ with $v$ 
the typical matter velocity, due to infall or peculiar motions.  With an 
impulse approximation (15)
becomes $z\sim mH_0 b^{-1}v\sim\ep^3$ since $v\sim\phi^{1/2}\sim\ep$ for
bound systems.  From this order of magnitude parametrization it is clear 
that to obtain a frequency shift (or equivalently temperature shift in 
the CMB) large enough to be observable (say $>10^{-6}$) one needs a rapid 
variation in the potential, more rapid than the gravitational time scale, 
such as from relativistic cosmic strings or black holes (but recall that 
the metric (1) was  derived for nonrelativistic, weak field 
perturbations).  Thus, 
within the context of a dynamical 
Sachs-Wolfe effect, the post-Newtonian formalism provides a significant 
but unobservable difference. 
\smallskip 
\leftline{\bf 3.3 Fourier decomposition} 
\noindent 
For a more diffuse density distribution, e.g. a linear density field, it 
is convenient to Fourier decompose the gravitational potential 
into modes corresponding to a characteristic fluctuation wavelength, or 
density inhomogeneity length or mass scale.  In fact, working in Fourier 
space allows simplification of many expressions.  For example, to 
obtain the Laplacian $\nabla^2\phi$ for arbitrary $p$, just transform (3), 
multiply by $q^2$, and perform the inverse transform.  One finds that 
for general $p$, 
(11) is altered by a term $(a/a_0)^p$ multiplying $f(\vx)$ and a sum 
inside the integral of polynomials up to order $p$ involving 
$|\vx-\vy|^2$ and $C_0$, multiplying the exponential. 

Having seen that 
the variation of the gravitational potential over the 
photon propagation gives an insufficiently large effect to be readily 
observable, we are led to consider the effect of the potential at the 
endpoints of the path, i.e. at emitter and observer, on the photon energy. 
This gravitational redshift is simply $z=\phi_e-\phi_o$.  Whereas in 
the isolated mass cases of \S3.1, 3.2 we could ignore the initial condition 
term in (3), either by saying it provided a term in the redshift that 
just added independently to the propagation effect, or by realizing 
that its contribution was exponentially suppressed far from the mass, 
now when dealing with a density field and an endpoint effect we must 
treat it more carefully.  Either we can ignore it by pushing $\eta_0\to0$, 
or include it in the calculation.  We choose the latter. 

We concentrate on the last scattering surface of the microwave background 
radiation and the density perturbations $\Dl_k$ there, writing the 
wavenumber from now on as $k$ instead 
of $q$ to agree with the usual notation in the 
literature.  In 
the linear density perturbation regime $\Dl$ grows linearly with scale 
factor $a$, corresponding to our case $p=1$.  The mean square gravitational 
redshift is proportional to the sum over Fourier modes of the square of 
the gravitational potential, assuming random phases between modes: $\langle 
z^2\rangle \sim \int d^3k\,|\phi_k|^2$.  The magnitude of the temperature 
anisotropies $(\Dl T/T)^2=\langle z^2\rangle$ is usually 
characterized by the zero separation correlation function $C(0)$ but our 
case is slightly different.  For lines of sight separated by some large 
angle $\psi$ the two point correlation function $C(\psi)=\langle z(\vec 
0)\,z(\vec\psi)\rangle$ does not vanish because of the coherence 
introduced by the $\langle\phi_o^2\rangle$ term.  Rather than have 
$(\Dl T/T)^2=2[C(0)-C(\psi)]\to 2C(0)$ for $\psi$ large, 
one has $(\Dl T/T)^2\to\langle \phi_e^2\rangle$, i.e. $\langle\phi_e^2 
\rangle$ plays the role of the zero lag correlation in setting the 
magnitude.  Alternately one can say that $\phi_o$ offers only a 
constant, isotropic shift so one can ignore it and consider only 
$z=\phi_e$. 

This quantity can be written in terms of the density 
perturbation power spectrum by taking the Fourier transform of (8b), 
but first we must evaluate the initial condition term of (3) and add 
it in.  It is 
$$\phi_k^{(ic)}(\eta)=[a(\eta_0)/a(\eta)]\,\phi_k(\eta_0)\,e^{-k^2 
C_0}.\eqno(16)$$ 
Note the exponential suppression as $\eta$ grows larger (later) than 
$\eta_0$. 
At this point we see that we really have two inputs to 
specify in (3), the initial density field $\Dl(\eta_0)$ and the initial 
potential $\phi(\eta_0)$.  Because of the form of the full relativistic 
equation (2), both must be given.  This is not unexpected though, because 
in JLW2 it was pointed out that (2) was essentially analogous to an 
inhomogeneous 
diffusion equation with time dependent parameters.  It was found that 
the potential corresponded to the temperature in that sort of problem, 
and the density contrast to a heat source term.  From that physical 
situation, however, we know that we must generally specify the 
initial distribution of both the temperature and the 
heat sources.  It is only in the 
late time limit that the source completely determines the diffusive 
variable; in our case this corresponds (if compactness holds as 
well) to the Newtonian limit and the 
Poisson-Newton equation. 

For the initial condition constant $\phi_k(\eta_0)$ we 
use the ansatz $\phi_k(\eta_0)=-4\pi Ak^{-2}f_k$, so $A$ gives the 
deviation from the Poisson equation ($A=1$) in the initial 
conditions.  
Adding (16) to the transform of (8b) gives 
in terms of the density perturbation power spectrum 
$$\eqalign{\langle z^2\rangle&=16\pi^2\int d^3k\,P_k k^{-4}{\cal C}_k,\cr 
{\cal C}_k&=\left[1-(1-A)(\eta_0/\eta)^2e^{-k^2C_0}-12k^{-2}\eta^{-2} 
(1-e^{-k^2C_0})\right]^2,\cr}\eqno(17)$$ 
where $P_k=|\Dl_k|^2$ is the density power spectrum and $C_0=C(\eta_0, 
\eta)$ as given by (4). 

Note that ${\cal C}$ gives the overall correction factor to the usual 
Poisson relation (in Fourier space) between the gravitational potential 
$\phi$ entering the metric (1) 
and the energy density; in particular, ${\cal C}=A^2$ at $\eta=\eta_0$. 
It can also be obtained directly from transforming the $p=1$ version 
of the Laplacian (11).  Thus ${\cal C}$ can be viewed alternately as 
giving the 
post-Newtonian adaptation of the endpoint Sachs-Wolfe effect, or as an 
effective alteration to the density power spectrum $P_k$, i.e.~an effective 
transfer function.  It is included in any (e.g.~numerical) treatment 
using the full equation (2). 

Depending on the initial conditions it has the possibility of 
either enhancing or suppressing the low Fourier modes, i.e. 
increasing or decreasing the large scale power in the intrinsic power 
spectrum $P_k$.  This, for 
example, would cause a smaller (larger) overall normalization factor to be 
needed to match the large angle COBE microwave background anisotropy 
measurements, and hence also decrease (increase) the resultant predicted 
small scale power, thus ameliorating (exacerbating) the difficulties of 
the cold dark matter model. 

The correction factor 
${\cal C}_k$ is plotted in Figure 1.  In the late time limit, 
$\eta\gg\eta_0$, the results are independent of $\eta/\eta_0$ and $A$.  
From the expression in (17) or the Figure, three regimes in 
perturbation wavelength 
can be identified.  When $k\gg C_0^{-1/2}$, the factor is close to one, 
i.e. this is the late time, compact inhomogeneity Newtonian limit. 
For $\sqrt{12}\eta^{-1}\ll k\ll C_0^{-1/2}$, the factor begins to decline 
predominately due to the influence of the exponential, and when $k\ll 
\sqrt{12}\eta^{-1}$, the $(k\eta)^{-2}$ term becomes important as well, 
leading to ${\cal C}_k\to A^2(\eta_0/\eta)^4$.  
From Figure 1, the deviation from the usual $P_kk^{-4}$ behavior of the 
Sachs-Wolfe effect becomes noticeable below $k\eta=10-40$, depending 
on the value of $C_0$.  Converting to wavelengths by 
$$k\eta=4\pi\eta/(H_0\lambda_0)=120\,(\lambda_0/10h^{-1}Mpc)^{-1} 
[(1+z_e)/10^3]^{-1/2},\eqno(18)$$ 
with $\lambda_0$ the present day wavelength and $z_e$ the redshift 
of the last scattering surface, 
we see this corresponds to scales $\lambda_0>30-100\,h^{-1}Mpc$ or 
angular scales $\theta>(1/2)-2^\circ$, applicable to the COBE regime. 

\bigskip 
\leftline{\bf 4. Conclusion}
\medskip\noindent 
The pseudo-Newtonian gravitational potential appearing in the metric 
(1) that realistically approximates our universe, incorporating 
inhomogeneities in a FRW background, is related nontrivially to that 
density fluctuation distribution.  The Green function (or kernel or 
propagator, in field theory terms) generally has a different spatial 
extent, or compactness, than the Newtonian case which corresponds to 
its late time, near neighborhood limit.  In consequence, light 
propagation behavior for those regions of parameter space far from that 
limit differ from the Newtonian. 

In the dynamical Sachs-Wolfe effect of the redshift induced by 
inhomogeneities along the light path, the deviation was shown to be 
significant but the overall magnitude too small to be observationally 
interesting.  For the endpoint Sachs-Wolfe effect of gravitational 
potential fluctuations at the source, the deviation could be viewed as 
an effective transfer function modifying the density perturbation 
power spectrum, with potentially important consequences for the 
normalization of the primordial density spectrum and hence the 
predictions of that model on smaller scales as a galaxy formation 
scenario. 

\bigskip 
\noindent{\it Acknowledgments.} 
I thank Mark Jacobs, Jason Kendall, and Bob Wagoner for useful 
discussions,   and the referee, Ed  Bertschinger, for suggesting 
clarifications. 
\medskip
\leftline{\bf References} 
\smallskip\noindent 
\noindent Bardeen, J.M., 1980, PRD, 22, 1882 

\noindent Futamase, T., 1989, MNRAS, 237, 187 

\noindent Futamase, T., Schutz, B.F., PRD, 28, 2363

\noindent Jacobs, M.W., Linder, E.V., Wagoner, R.V., 1992, 
PRD, 45, R3292 (JLW1)

\noindent Jacobs, M.W., Linder, E.V., Wagoner, R.V., 1993, 
PRD, 48, 4623 (JLW2)

\noindent Kendall, J.S., 1993, unpublished MSc thesis, New Mexico State 
University 

\noindent Kodama, H., Sasaki, M., 1984, Prog. Th. Phys. Supp, 
78, 1 

\noindent Ma, C-P., Bertschinger, E., 1995, ApJ, 455, 7

\noindent Rees, M.J., Sciama, D.W., 1968, Nat, 217, 511 

\noindent Seljak, U., 1996, ApJ, 460, 549
\medskip 
{\leftline{\bf Figure caption}
\smallskip\noindent 
{\bf Fig.~1.} The Poisson correction factor ${\cal C}_k=|\phi_k|^2/
(16\pi^2k^{-4}P_k)$ is plotted vs.~$k\eta$ for various initial conditions 
$(A,\eta/\eta_0)$.  For small wavenumbers (large scales) a significant 
enhancement or suppression can be achieved.  The late time curve 
$\eta\gg\eta_0$ is independent of $A$.} 
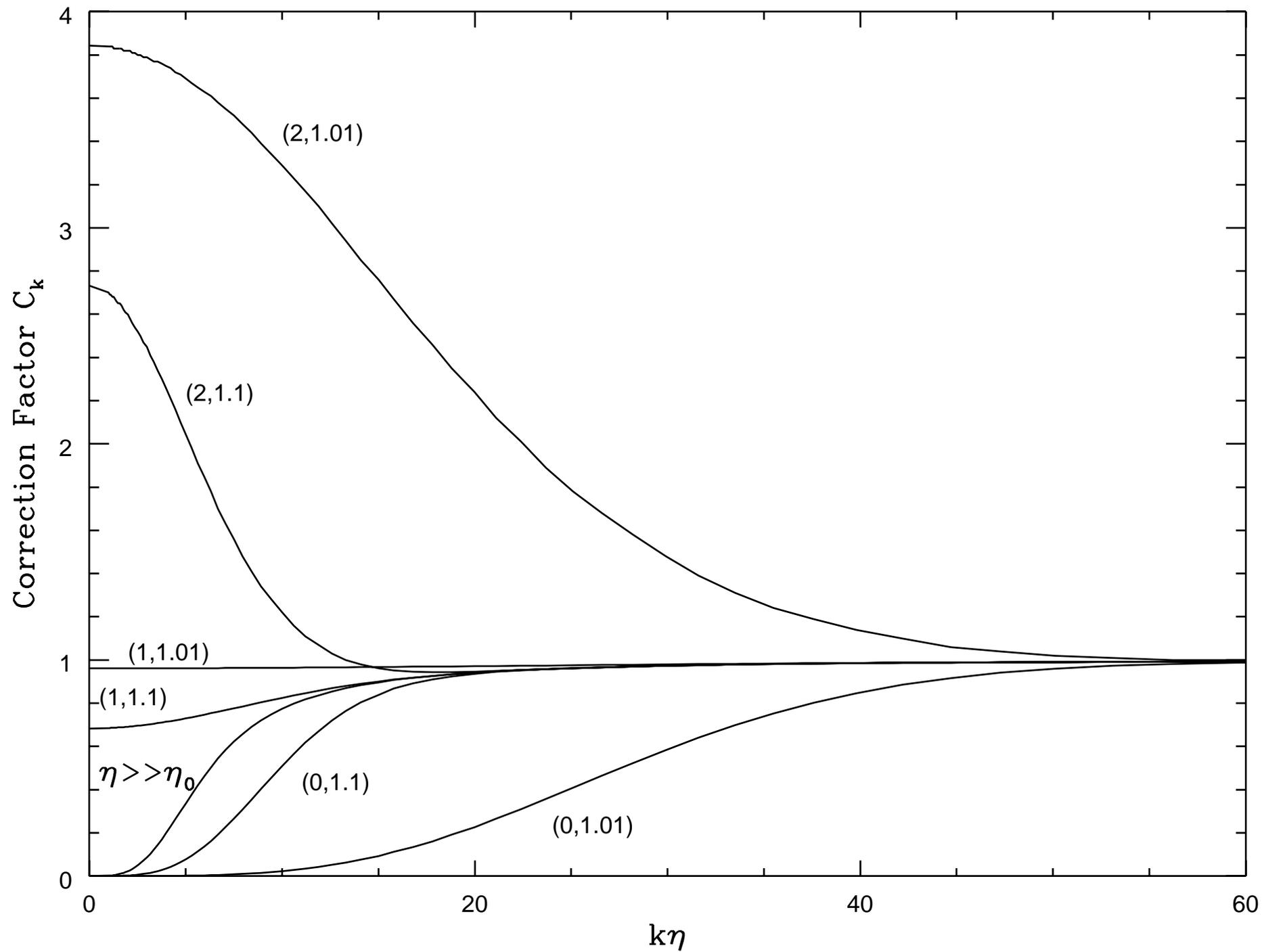
\bye